\documentclass{moriond}
\usepackage{subfig}

\def\be{\begin{equation}}
\def\ee{\end{equation}}
\def\bea{\begin{eqnarray}}
\def\eea{\end{eqnarray}}

\newcommand{\Photo}{}

\begin{document}
\vspace*{4cm}
\title{Top-Quark Properties at the LHC}

\author{ K. Beernaert on behalf of the ATLAS and CMS collaborations}

\address{DESY CMS, Notkestra{\ss}e 85,\\
22607 Hamburg, Germany}

\maketitle\abstracts{A review of recent measurements of top-quark properties is presented. Inclusive and differential top-quark 
pair charge asymmetry measurements using the full Run I dataset are found to be in agreement with the standard model (SM) predictions. 
Results of spin correlation in top-quark pairs are presented 
and interpreted in terms of the SM predicted values and new physics models. 
Limits are set on flavour-changing neutral currents (FCNC), in particular with a Higgs boson in the 
final state.}

\section{Introduction}
\label{sec:Introduction}
Due to its short lifetime, the top quark decays before it can form bound states and before its spin decorrelates. As a 
consequence we can study "bare" quark properties. The top quark has a mass of approximately $173$ GeV and may play a significant 
role in electro-weak symmetry breaking due to its large coupling to the Higgs boson. 
Measurements of the top-quark properties with increasing levels of precision test the SM and open the possibility to 
probe new physics. The data used for the studies presented here were collected in pp collisions in 2011 and 2012 at centre-of-mass energies of 7 and 
8 TeV at the Large Hadron Collider (LHC) by the ATLAS~\cite{1748-0221-3-08-S08003} and CMS~\cite{Chatrchyan:2008aa} detectors.

\section{Production Asymmetries}
\label{sec:Production Asymmetries}
The Tevatron forward-backward asymmetry measurements have initially shown some tension with the SM predictions~\cite{D0,CDF}. 
At the LHC, with a symmetric initial 
state, a charge asymmetry is measured, given by:

\begin{equation}
  A_{C} = \frac{N(\Delta \left| y \right| > 0) - N(\Delta \left| y \right| < 0)}{N(\Delta \left| y \right| > 0) + N(\Delta \left| y \right| < 0\
)}
\label{eq.ChargeAsym}
\end{equation}
\noindent
The charge asymmetry can be set up using the rapidity $y$ of the top quarks 
($\Delta \left| y \right|\,=\, \left| y_{t} \right| - \left| y_{\overline{t}} \right|$) or with the pseudo-rapidity $\eta$ 
of the leptons in the dilepton channel (replacing $\Delta \left| y \right|$ with 
$\Delta \left| \eta_{l} \right|\,=\,\left|\eta_{l^{+}}\right| - \left|\eta_{l^{-}} \right|$). In the SM, the top-quark 
production asymmetries are due to NLO QCD interference effects. Measurements of $A_{FB}$ at the Tevatron and $A_{C}$ at 
the LHC have complementary sensitivity to new physics models~\cite{PhysRevD.84.115013,AguilarSaavedra:2011ug}. ATLAS presents two 
results in the lepton+jets channel at $\sqrt{s}\,=$ 8 TeV. In one analysis~\cite{Phys.Lett.B756}, a minimum $t\overline{t}$ 
invariant mass of 0.75~TeV is imposed, and the final state is selected by looking for a resolved leptonic top-quark 
decay and a hadronic decay which is reconstructed as a large $R$-jet with substructure where $R = \sqrt{(\Delta\eta)^{2} + 
(\Delta\phi)^{2}}$. In the other presented analysis~\cite{Eur.Phys.J.C76}, three signal regions are used based on the 
b-tag multiplicity. Full Bayesian unfolding is used to bring the distributions back to parton level. The inclusive 
charge asymmetry is measured as $A_C = [4.2\, \pm\, 3.2\,\mathrm{(stat.+syst.)}]\,\%$~\cite{Phys.Lett.B756} and 
$A_C = [0.9\, \pm\, 0.5\,\mathrm{(stat.+syst.)}]\,\%$~\cite{Eur.Phys.J.C76}. 
CMS also presents an analysis in the lepton+jets channel~\cite{Phys.Rev.D93} at $\sqrt{s}\,=$~8~TeV. 
The charge asymmetry is measured with a template fit using symmetric and asymmetric templates of 
$\Upsilon_{t\overline{t}} = \tanh{\Delta \left| y \right|_{t\overline{t}}}$. The fit parameter $\alpha$ 
represents the relative contribution of the symmetric and asymmetric templates. This results in 
$A_C = [0.33\, \pm\, 0.42\,\mathrm{(stat.+syst.)}]\,\%$. CMS presents a result in the dilepton channel~\cite{arxiv:1603.06221},
where the asymmetry is determined using the final-state leptons and reconstructed top quarks, leading to 
$A_C = [1.1\, \pm\, 1.3\,\mathrm{(stat.+syst.)}]\,\%$ based on the top quarks and 
$A^{\mathrm{lep}}_C = [0.3\, \pm\, 0.7\,\mathrm{(stat.+syst.)}]\,\%$ based on the leptons. 
The results are observed to be consistent with the SM, as seen 
from the summary in Fig.~\ref{fig:Summary_Asym}. 
In addition, all analyses provide differential measurements of the 
charge asymmetry in a variety of variables, e.g. invariant mass $m(t\overline{t})$, transverse 
momentum $p_{T}(t\overline{t})$ and velocity $\beta(t\overline{t})$ of the top-quark pair. 

\section{Spin Correlations}
\label{sec:Spin Correlations}
In top-quark pair production, the SM predicts the spins of the top- and antitop-quark to be correlated. The spin information of the top quark can be accessed 
using the decay products. ATLAS presents an analysis~\cite{Phys.Rev.D93012002} making use of the following double differential 
cross section:

\begin{equation}
 \frac{1}{N}\frac{d^{2}N}{d\cos\theta_{1}dcos\theta_{2}} = \frac{1}{4}(1+B_{1}\cos\theta_{1} + B_{2}\cos\theta_{2} -C_{\mathrm{hel}}\cos\theta_{1}\cos\theta_{2})
\end{equation}
where $\theta$ is the angle between the lepton direction in the top (anti-)quark parent rest frame and the top (anti-)quark 
parent in the $t\overline{t}$ rest frame. $B_{1,2}$ are proportional to the top-quark polarisation and are considered to be zero. 
Using this equation, the spin correlation strength $A_\mathrm{hel}$ can be directly extracted from 
$C_\mathrm{hel} = -A_\mathrm{hel}\alpha_1 \alpha_2$. ATLAS presents an analysis at $\sqrt{s}\,=$ 7~TeV~\cite{Phys.Rev.D93012002} in the 
dilepton channel that 
results in $A_\mathrm{hel}\,=\,0.315\,\pm\,0.061\,\mathrm{(stat.)}\,\pm\,0.049\,\mathrm{(syst.)}$. 
CMS presents an analysis at $\sqrt{s}\,=$ 8~TeV~\cite{Phys.Rev.D93052007} 
in the dilepton 
channel where several asymmetry variables are used to perform a direct measurement of the spin correlation strength and the 
top-quark polarization. 
A measurement of the 
spin correlation strength can be interpreted in terms of several BSM models. 
As an example, the CMS measurement has been used to set limits on top-quark chromomagnetic couplings~\cite{Bernreuther} of 
$-0.053\,<\, \mathrm{Re}(\mu_t)\,<0.026$ for the chromomagnetic dipole moment (CMDM) and $-0.068\,<\,\mathrm{Im}(d_t)\,<\,0.067$ for the 
chromo-electric dipole moment (CEDM) both at the 95 \% confidence level (CL). ATLAS presents an analysis at $\sqrt{s}\,=$ 8~TeV~\cite{PRL114} where 
the spin correlation measurement is interpreted in terms of a Minimal Super-Symmetric Model (MSSM) where stop squarks decay 
100 \% into a top quark and a neutralino with the stop squark mass very close to the top-quark mass. Stop squark masses 
between the top-quark mass and 191~GeV are excluded at the 95 \% CL. CMS presents an analysis at 
$\sqrt{s}\,=$ 8~TeV~\cite{arxiv:1511.06170} in the lepton + jets channel where a matrix element method is used to set up 
a variable sensitive to spin correlation. Using matrix elements for $t\overline{t}$ production and decay using SM 
spin correlations and zero spin correlations, the likelihood ratio of these two hypotheses is used to perform a template 
fit and a hypothesis testing procedure. A SM fraction of 
$f^{SM}\,=\,0.72\,\pm\,0.08\,\mathrm{(stat.)}\,{}^{+0.15}_{-0.13}\,\mathrm{(syst.)}$ is measured.

\section{Flavour-changing Neutral Currents}
\label{sec:Flavour-changing Neutral Currents}
\begin{figure}
\subfloat[a][Summary of top charge asymmetry]{\includegraphics[width=0.5\textwidth]{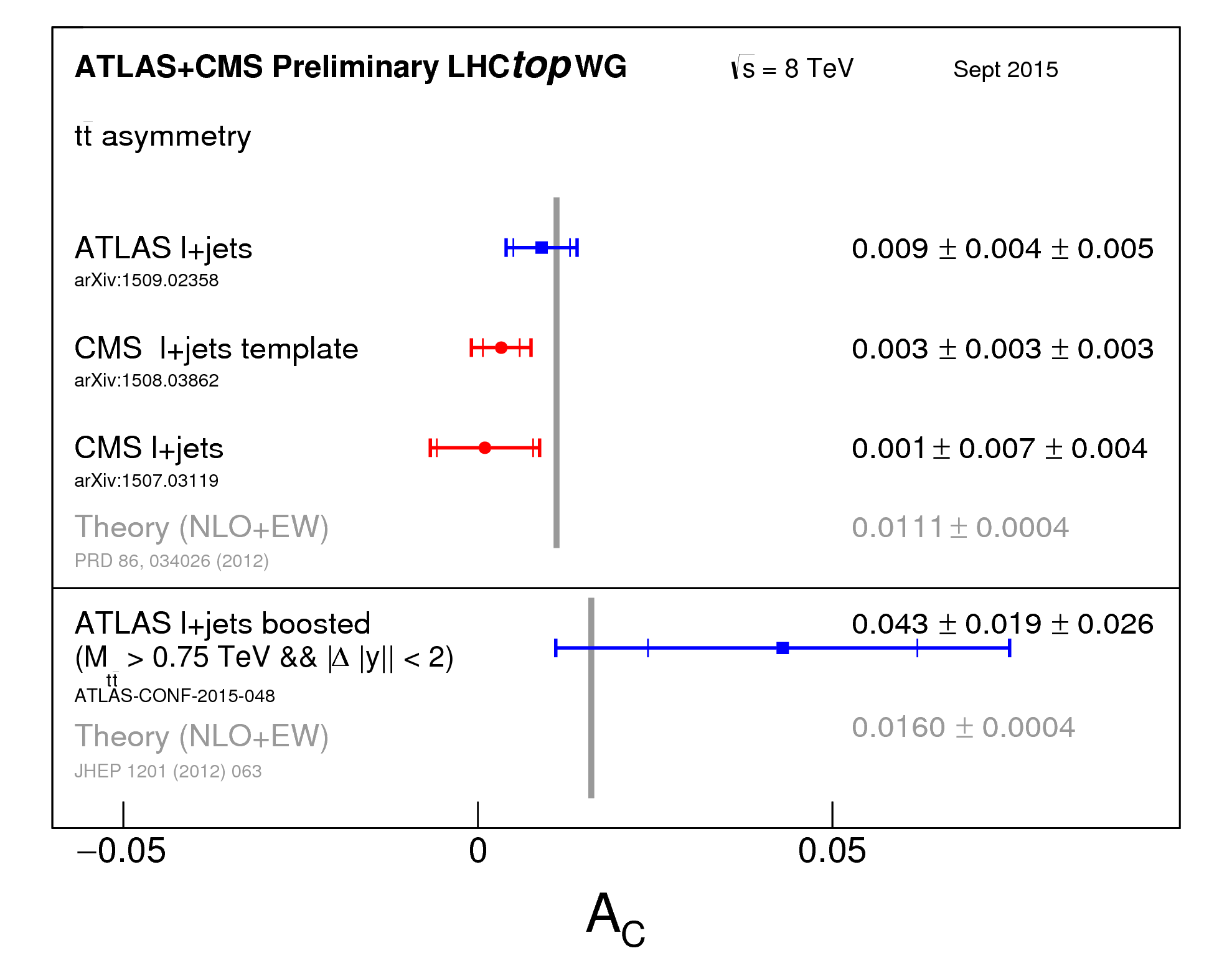}\label{fig:Summary_Asym}}
\subfloat[b][Summary of FCNC limits in CMS]{\includegraphics[width=0.5\textwidth]{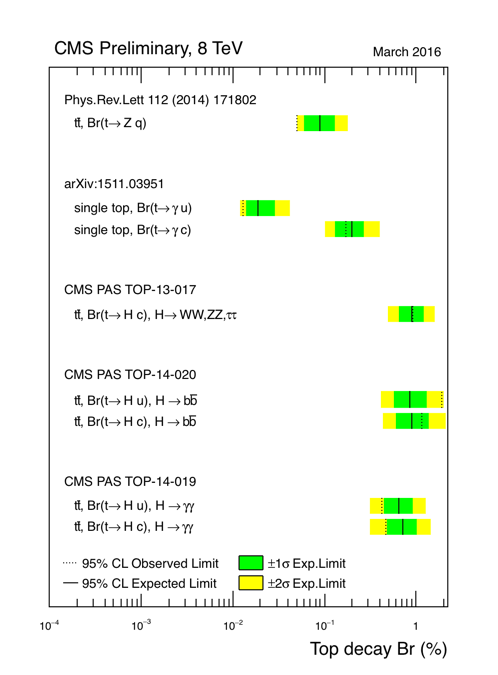}\label{fig:SummaryFCNC_CMS}}

\subfloat[c][Summary of FCNC limits in ATLAS $t\rightarrow cX$]{\includegraphics[width=0.5\textwidth]{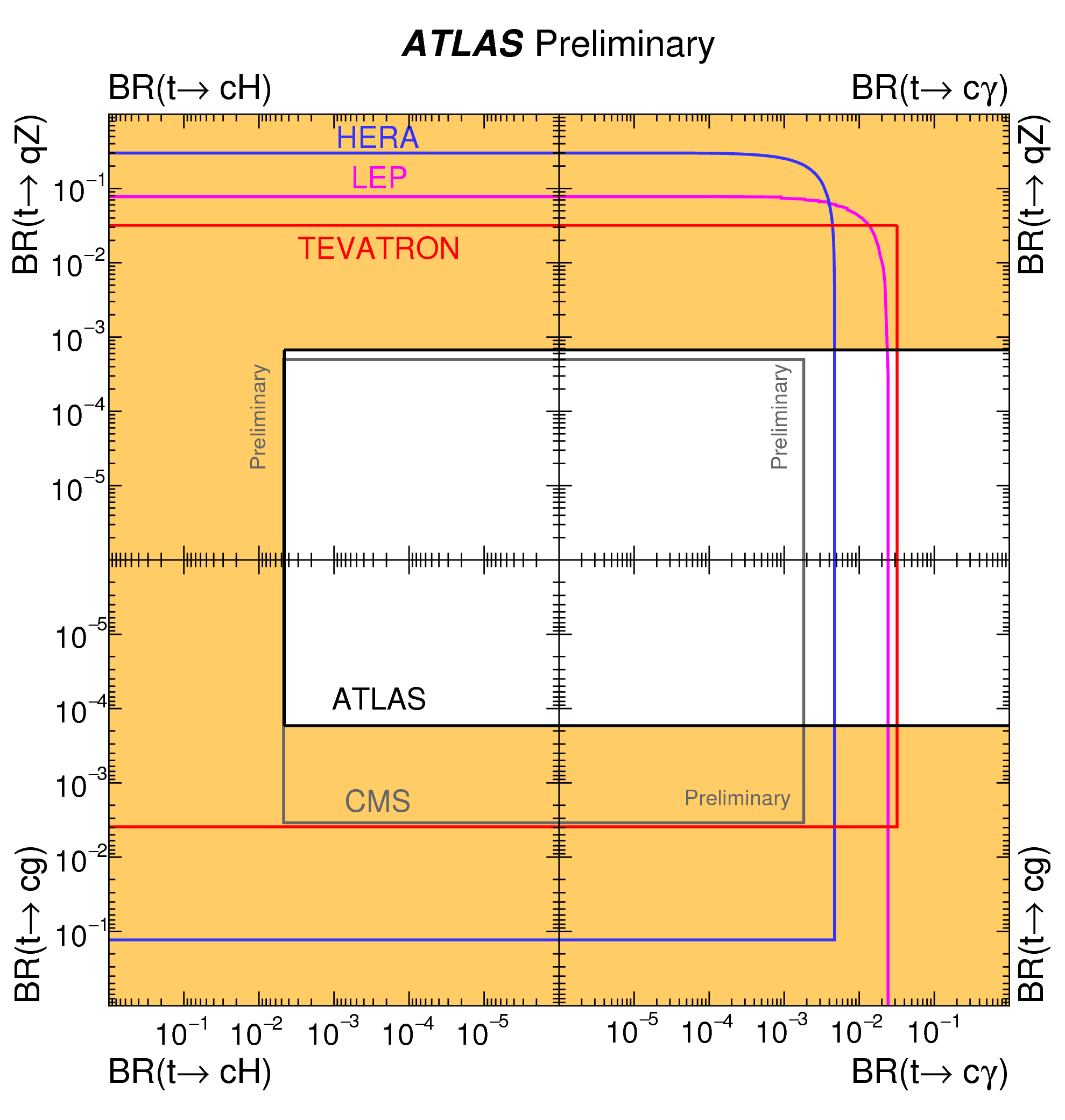}\label{fig:SummaryFCNC_ATLAStcX}}
\subfloat[d][Summary of FCNC limits in ATLAS $t \rightarrow uX$]{\includegraphics[width=0.5\textwidth]{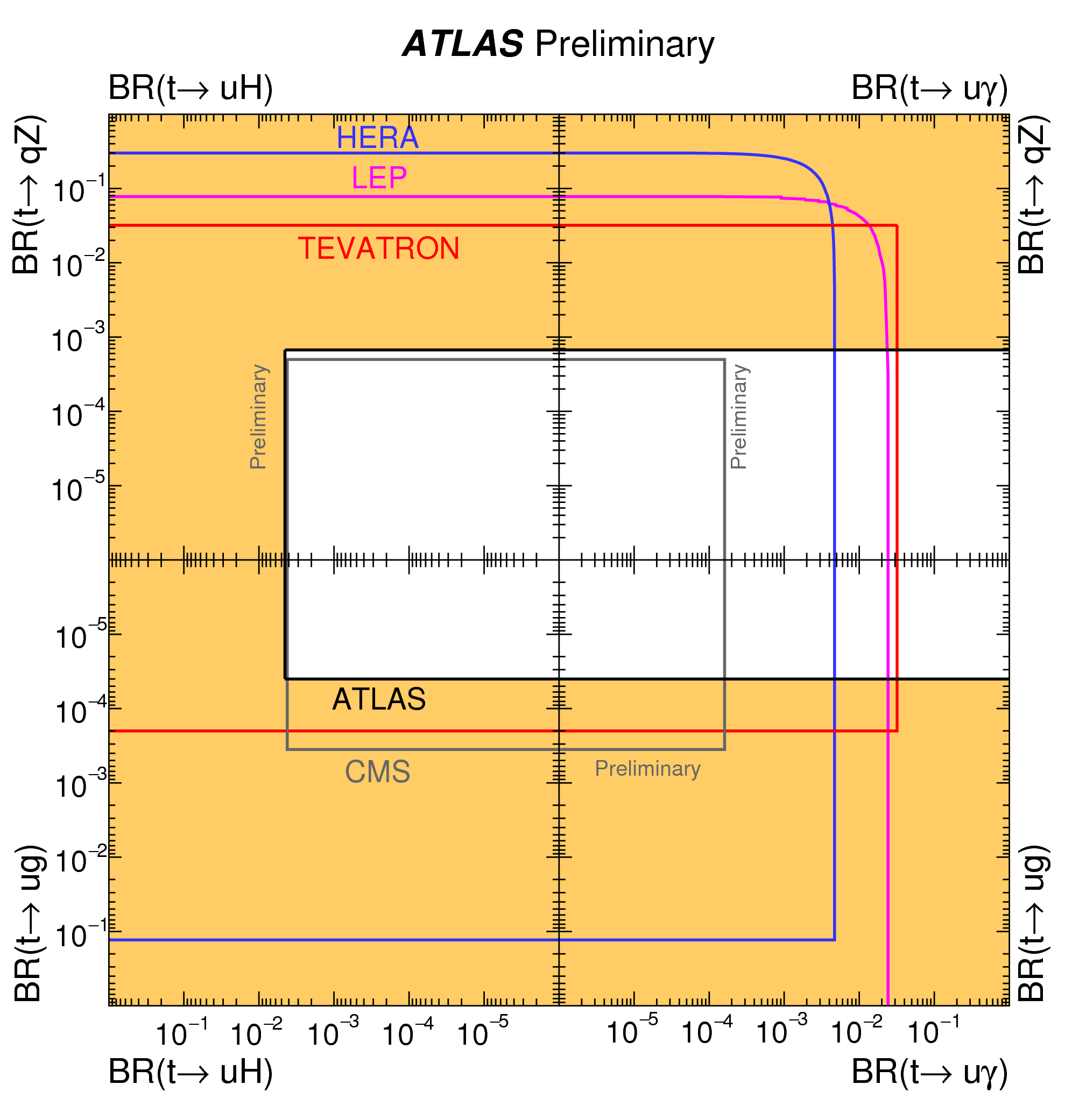}\label{fig:SummaryFCNC_ATLAStuX}}
\caption{Summary of top charge asymmetry measurements at $\sqrt{s}\,=\,8$~TeV at ATLAS and CMS in Fig.~\ref{fig:Summary_Asym}. Summary of the limits on FCNC for CMS 
in Fig.~\ref{fig:SummaryFCNC_CMS} and ATLAS in Fig.~\ref{fig:SummaryFCNC_ATLAStcX}-\ref{fig:SummaryFCNC_ATLAStuX}. In the ATLAS summary plots, the orange exclusion 
region narrows down to smaller branching ratios at the centre of the plot, while in the CMS summary plot the excluded region is on the right of 
the observed branching ratio limits.}
\label{fig.SummaryFCNC}
\end{figure}

In the SM, FCNC are suppressed at tree-level due to the GIM mechanism. This leads to very 
small branching ratios of $t \rightarrow u/c + X$ with $X = g, \gamma, Z, H$ of $O(10^{-12} - 10^{-17})$. Several 
BSM models, such as MSSM, 2HDM, predict enhanced couplings for FCNC with branching ratios as high as $10^{-5}$. With the 
discovery of the Higgs boson, FCNC can now also be studied in $t\overline{t}$ where one of the top quarks decays
as $t \rightarrow u/c + H$. A clear overview of the predictions can be found in the reviews~\cite{rev1,rev2}. CMS presents 
three analyses in this channel. One analysis~\cite{CMS-PAS-TOP-14-020} makes use of the high branching fraction of 
$H \rightarrow b\overline{b}$ to look for $t\overline{t}$ production with decays of 
$t \rightarrow Hq \rightarrow b\overline{b}q$ in 
one leg and $t \rightarrow Wb \rightarrow l\nu b$ in the other, obtaining an observed limit of 
$B(t \rightarrow Hc) < 1.16 \%$ and $B(t \rightarrow Hu) < 1.92 \%$ at 95 \% CL. Using the cleaner Higgs decay channel 
$H \rightarrow \gamma\gamma$ and looking for top-quark pairs with $t \rightarrow Hq \rightarrow \gamma\gamma q$ and 
$t \rightarrow Wb \rightarrow l\nu b\, \mathrm{or}\, q\overline{q}b$, observed limits are set of $B(t \rightarrow Hc) < 0.47 \%$ 
and $B(t \rightarrow Hu) < 0.42 \%$ at 95 \% CL~\cite{CMS-PAS-TOP-14-019}. Finally, using $t\overline{t}$ events where 
$t \rightarrow Hq \rightarrow ZZq\, \mathrm{or}\, WWq$ and $t \rightarrow Wq \rightarrow l\nu b$, CMS reports an 
observed limit of $B(t \rightarrow Hc) < 0.93 \%$ at 95 \% CL~\cite{CMS-PAS-TOP-13-017}. ATLAS presents an analysis at $\sqrt{s}\,=$~8~TeV~\cite{JHEP12_061} 
searching for FCNC in the channel $t \rightarrow Hq \rightarrow b\overline{b}q$ and $t \rightarrow Wb \rightarrow l\nu b$. 
Several signal categories are used based on jet and b-tag multiplicity. An observed limit of $B(t \rightarrow Hc) < 0.56 \%$ and 
$B(t \rightarrow Hu) < 0.61 \%$ is reported at the 95 \% CL. In addition, a re-interpretation of previous $t\overline{t}H$ searches 
is performed in this analysis and 
combined limits are presented. Summaries of the observed limits on FCNC are shown in Fig.~\ref{fig:SummaryFCNC_CMS}-\ref{fig:SummaryFCNC_ATLAStuX}.

\section{CP Violation}
\label{sec:CP Violation}
A first search for CP violation in the $t\overline{t}$ sector has been pursued by CMS in the l+jets channel by inspecting T-odd observables~\cite{CMS-PAS-TOP-16-001}. 
The observables are defined as $O_{2}\,\propto\,(\overrightarrow{p}_{b} + \overrightarrow{p}_{\overline{b}})\cdotp (\overrightarrow{p}_{l} \times \overrightarrow{p}_{j1})$, 
$O_{3}\,\propto\,Q_{l}\overrightarrow{p}_{b}\cdotp(\overrightarrow{p}_{l} \times \overrightarrow{p}_{j1})$, 
$O_{4}\,\propto\,Q_{l}(\overrightarrow{p}_{b} - \overrightarrow{p}_{\overline{b}})\cdotp(\overrightarrow{p}_{l} \times \overrightarrow{p}_{j1})$ and 
$O_{7}\,\propto\,(\overrightarrow{p}_{b} - \overrightarrow{p}_{\overline{b}})_{z}(\overrightarrow{p}_{b} \times \overrightarrow{p}_{\overline{b}})_{z}$.
The statistically limited results are found to be in agreement with no CP violation in $t\overline{t}$ production and decay,
and the following values have been measured for $A'_{CP}$: $O_2=+0.27\,\pm\,0.41\,\mathrm{(stat.)}\,\pm\,0.01\,\mathrm{(syst.)}$, 
$O_3\,=\,-0.71\,\pm0.41\,\mathrm{(stat.)}\,\pm\,0.03\,\mathrm{(syst.)}$, $O_4\,=\,-0.38\,\pm\,0.41\,\mathrm{(stat.)}\,\pm\,0.03\,\mathrm{(syst.)}$, 
$O_7\,=\,-0.06\,\pm\,0.41\,\mathrm{(stat.)}\,\pm\,0.01\,\mathrm{(syst.)}$.

\section{Conclusions}
\label{sec:Conclusions}
Top-quark properties measurements at the LHC provide precision tests of the SM. Limits have been set on FCNC and precision 
measurements of spin correlations and the charge asymmetry are consistent with the SM.

\end{document}